\documentclass[doublecol]{epl2} 
\usepackage{graphicx,amsmath,amssymb}
\usepackage{hyperref}

\title{Thermal Transport Imaging in the Quantum Hall Edge Channel}

\author{J. N. Moore\inst{1} \and A. Kamiyama\inst{1} \and T. Mano\inst{2} \and G. Yusa\inst{1}}

\institute{                    
  \inst{1}Department of Physics, Tohoku University - Sendai 980-8578, Japan\\
  \inst{2}National Institute for Materials Science - Tsukuba, Ibaraki 305-0047, Japan
}

\abstract{Research focused on heat transport in the quantum Hall (QH) edge channel has successfully addressed fundamental theoretical questions surrounding the QH physics. However, the picture of the edge channel is complicated by the phenomenon of energy dissipation out of the edge, and theories treating this dissipation are lacking. More experimental data is also needed to determine the coupling mechanism by which energy leaves the edge channel. We developed a method to map the heat transport in the QH edge to study the dissipation of heat. We locally heated the QH edge and locally detected the temperature increase while continuously varying the distance between heater and thermometer. We thereby obtained the thermal decay length of the edge state.}

\begin{document}

\maketitle

\section{Introduction}
Heat transport in the quantum Hall (QH) edge channel has garnered much recent attention. This interest began with the prediction and subsequent verification of a charge-neutral mode in certain fractional QH edge states, which carries heat counter to the direction of the charge mode \cite{kane95,bid}. The discovery of the charge-neutral mode was followed by further effort to understand the complex structure of the QH edge state. For this, charge transport measurements alone do not provide sufficient detail; but measurements of heat transport and thermal conductance in the edge have generated significant progress in identifying the composition of compound edge channels and revealing the process of equilibration between edge modes \cite{lesueur,venkatachalam,srivastav,lebreton}. Thermal conductance has also been critical to testing the order of the $\nu=5/2$ QH state and providing evidence for the non-Abelian statistics of its quasiparticles \cite{banerjee}. $\nu=\frac{h}{eB}n_{\text{e}}$ is the Landau-level filling factor, where $h$ is the Planck constant, $e$ is the elementary charge, $B$ is the perpendicular magnetic field and $n_{\text{e}}$ is the 2D electron density.

Theoretical proposals have been made to use the chiral heat transport of the QH edge to produce several types of mesoscopic quantum heat devices such as a refrigerator, heat engine, heat rectifier and heat diode \cite{sanchez2019,sanchez2015,vannucci,sanchez2015new}. These devices should operate very efficiently if heat travels ballistically in the edge. An early theoretical assumption of the edge channel was that, just as electrons do not leave the edge channel due to an absence of available states in the bulk, heat will also remain inside the edge channel \cite{kane97}. However, a thermal decay length of  $\sim20$ $\mu$m was measured in the $\nu=1$ \cite{granger} and $\nu=2$ QH states \cite{nam}. This is significantly shorter than the decay length of charge transport, as can be inferred from $\sim1$-mm-long samples which display vanishingly small longitudinal resistance. The cause of heat dissipation from the QH edge is unknown, but is likely due in part to coupling with the two-dimensional (2D) QH bulk; heat propagation through the bulk was detected by experiments using three different methodologies \cite{venkatachalam,altimiras,inoue}. Mechanisms considered were 1) long-rang Coulomb interaction with localized bulk states, 2) coupling with bulk spin states 3) and coupling with the lattice.

We developed an optical heating technique to investigate QH edge heat transport and heat dissipation. Illuminating a sample with a focused laser beam  allows for the local heating of electrons at any arbitrary position, which can be changed \textit{in situ}. Laser heating also offers the potential of modulating the heating power to measure thermal relaxation times as short as $\sim1$ ps \cite{aamir}. We combined optical heating with local thermometry of the QH edge to acquire spatial information about heat transport. We succeeded in measuring the thermal decay length of the edge channel by measuring the temperature response as a function of the distance between the heater and thermometer. When temperature response is spatially resolved in sufficient detail, it can inform the creation of new models which incorporate energy dissipation out of the edge \cite{steinberg}. With this goal, we performed the first continuous real-space imaging of heat transport in the QH edge. 

\section{Experiment}
The 2D electron system (2DES) was contained in a Al$_{0.3}$Ga$_{0.7}$As/GaAs/Al$_{0.3}$Ga$_{0.7}$As 15-nm quantum well (QW) grown on a Si-doped (100) GaAs substrate \cite{moore2016,moore2017,moore2018}. The conducting substrate was insulated from the 2DES by a GaAs/AlAs superlattice; this allowed us to use the substrate as a global back gate of the 2DES. We modulation doped the QW by a unique superlattice $\delta$-doping structure designed to suppress photoconductivity in channels parallel to the QW so that our transport measurements will primarily detect conduction in the 2DES (see Superlattice Doping Structure in Supplementary Material for details). Figure \ref{fig.1} shows the band energy and 3D electron density $n_{\text{3D}}$ of the sample self-consistently simulated as a function of distance $z$ from the sample surface. Residual accepters which were unintentionally doped during the wafer growth are ignored in the calculation. The experimentally measured $n_{\text{e}}$ and mobility of the QW in this structure were $1.7\times10^{11}$ cm$^{-2}$ and $0.6\times10^{6}$ cm$^{2}$/V s respectively.

\begin{figure}
\onefigure[scale=0.45]{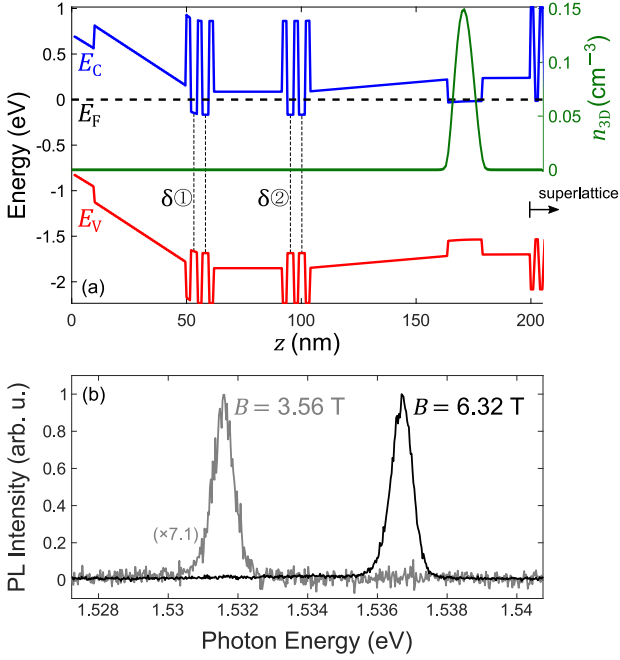}
\caption{(a) The conduction band energy $E_{\text{C}}$, heavy hole valence band energy $E_{\text{V}}$, Fermi energy $E_{\text{F}}$, and 3D electron density $n_{\text{3D}}$ self-consistently simulated at the gamma point and plotted as a function of distance $z$ from the surface at a location $1$ $\mu$m from the etched mesa edge. The vertical dashed lines indicated as $\delta\textcircled{1}$ ($\delta\textcircled{2}$) are $\delta$-dopes having Si doping concentration $6\times10^{11}$ cm$^{-2}$ ($3\times10^{11}$ cm$^{-2}$). (b) The $\sigma^-$-polarized photoluminescence spectrum, corresponding to recombination in the bottom Zeeman level, collected at two different magnetic field strengths $B$.}
\label{fig.1}
\end{figure}

We conducted the following experiments at a lattice temperature $T_{\text{L}}$ of 1.7 K, unless otherwise stated, while the 2DES was tuned to the $\nu=2$ QH state by adjusting the global back gate voltage and $B$ to the center of the QH resistance plateau. To create local heating, we illuminated the sample by a linearly polarized 785-nm diffraction-limited continuous wave laser spot of diameter $\sim1$ $\mu$m\cite{moore2016,moore2017,moore2018,kamiyama}. The laser photon energy, $E_{\text{ph}}=1.579$ eV, was greater than the ground state energy of an electron-hole pair in the QW, $E_{\text{rec}}=1.532\sim1.537$ eV, which we measured in this sample as described below. $E_{\text{ph}}$ was also less than the bandgap of AlGaAs ($\sim1.9$ eV)\cite{lourenco}. Thus, photon energy was absorbed in the QW, but not in the neighboring AlGaAs. Photon energy was, however, absorbed by the GaAs substrate, which is separated from the QW by a 200-period GaAs/AlAs (2 nm/2 nm) superlattice. Some of the energy absorbed in the substrate was converted to heat which was diffused through the sample by phonons. We numerically simulated the resulting phonon temperature throughout the sample and found that it increased by only $\sim0.1$ $\mu$K (see 3D Simulation of Phonon Heating in Supplementary Material for details), which is due to the large phonon thermal conductivity and the freedom of phonons to diffuse heat in three dimensions. 

The QH edge, in contrast, conducts heat in only one dimension with a thermal conductivity $K_{\text{H}}=\nu_{\text{Q}}\frac{\pi^2}{3}\frac{k_{\text{B}}^2}{h}T$, where $\nu_{\text{Q}}=2$ is the number of downstream heat conducting edge modes, $T=1.7$ K is the edge temperature, and $k_{\text{B}}$ and $h$ are the Boltzmann and Planck constants respectively \cite{kane97}. Now we estimate the temperature increase of the QH edge, $\Delta$$T_{\text{edge}}$. We take the absorption coefficient of the QW to be $1.7\times10^{4}$ cm$^{-1}$ \cite{marquezini}, from which we calculate that $2.5\%$ of the incident optical power will be absorbed in the QW. Most of the absorbed energy is consumed in generating an electron-hole pair across the bandgap. The maximum energy available for heat generation is $E_{\text{ph}}-E_{\text{rec}}$. We measured $E_{\text{rec}}$ from the peak energy in the photoluminescence spectrum collected during the experiment (Fig. \ref{fig.1}b). We interpret this peak to correspond to recombination of singlet-state trions, which are a bound state of a hole with two electrons \cite{buhmann}, because as a function of $n_{\text{e}}$ this peak is continuous with the singlet trion peak identified at $\nu<1$ \cite{shields, yusa}. The maximum fraction of absorbed energy that can be converted to heat is therefore $(E_{\text{ph}}-E_{\text{rec}}$)/$E_{\text{ph}}=2.9\%$. Using the incident illumination of 3 nW in our experiment, we calculate $\Delta$$T_{\text{edge}}=830$ mK, which is an upper estimate assuming the stated heat conversion efficiency and neglecting any dissipation of heat out of the edge into the environment.

\begin{figure}
\onefigure[scale=0.45]{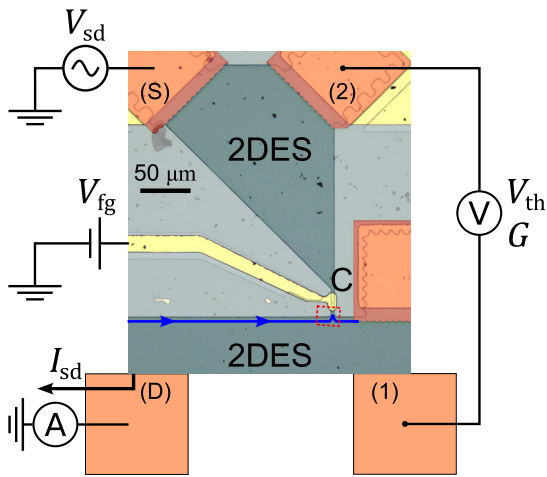}
\caption{Optical microscope image of the experimental device. C: 3.0 $\mu$m constriction in the 2DES covered by a gold gate. Red dotted outline: the region mapped in Fig. \ref{fig.3}. Blue solid line: The path of the QH edge when $B$ is oriented as indicated in Fig. \ref{fig.3}(b). Ohmic electrodes (colored orange): (S) voltage source, (D) current drain, (1) and (2) voltage probes.}
\label{fig.2}
\end{figure}

Following Ref. \cite{granger}, we used thermoelectric thermometry to locally detect the increase in the QH edge temperature. This method of thermometry measures the thermoelectric voltage $V_{\text{th}}$ that is generated across a gradient in electron temperature. Because $T_{\text{L}}$ does not increase significantly, the phonon drag contribution to $V_{\text{th}}$ can be ignored \cite{cantrell}. In our device, shown in Fig. \ref{fig.2}, $V_{\text{th}}$ appeared across a gated $3$-$\mu$m constriction, labeled ``C'', which we created by photolithography and etching. We generated a temperature gradient across this constriction by laser heating the 2DES on the lower side of the constriction. Across the constriction, $V_{\text{th}}=-\int_{T_{\text{1}}}^{T_{\text{2}}}\;S\text{d}T$, where $T_{\text{1}}$ and $T_{\text{2}}$ are the electron temperatures on the lower and upper side of the constriction respectively, and $S$ is the Seebeck coefficient depending on temperature and $n_{\text{e}}$. For small thermal gradients, $V_{\text{th}}$ is proportional to $T_{\text{1}}-T_{\text{2}}$, with $S$ determining the sensitivity of the thermometer.

Under the gated region of the constriction, $S$ is modified by the front gate voltage $V_{\text{fg}}$ and is determined by the Mott relation to exhibit peaks corresponding to the steps between plateaus in the constriction conductance \cite{appleyard,nam}. Past work at $T_{\text{L}}=1.65$ K used a peak in $S$ of $-20$ $\mu$V/K to achieve a thermometer sensitivity of $\sim2$ mK \cite{vanhouten}. Because of the incompressibility of the 2DES in the $\nu=2$ QH state, $S_{xx}=S_{xy}=0$ outside the gated region \cite{fletcher}. Therefore, although the laser heating creates a temperature gradient along the QH edge, no thermoelectric voltage is generated there. This allowed us to use Ohmic contacts (labeled (1) and (2)) connected to the 2DES on the lower and upper sides of the constriction to measure $V_{\text{th}}$ generated in the gated region alone. Note that under the illumination intensity of our experiment ($\sim3$ nW), we estimate the ratio of photoexcited carriers to native 2D electrons to be $\sim10$ to 1 million. This indicates that the photocarriers cannot significantly modify $S$ or independently generate photovoltage.

Simultaneously with the measurement of $V_{\text{th}}$, we measured the source-drain current $I_{\text{sd}}$ and four-terminal conductance $G$ through the constriction using a constant source-drain excitation voltage $V_{\text{sd}}$ of 10 $\mu$V. We measured $V_{\text{th}}$ and $G$ by lock-in detection with modulation of $V_{\text{sd}}$ and the laser power at 13.1 Hz and 5.94 kHz respectively. We performed all measurements at fixed phase and plotted the amplitude of the in-phase component, denoted by $\left|\cdot\right|$. We measured $V_{\text{th}}$ at a phase of $67^{\circ}$, where it's amplitude was maximum.

\section{Results and Discussion}
\begin{figure}
\onefigure[scale=0.31]{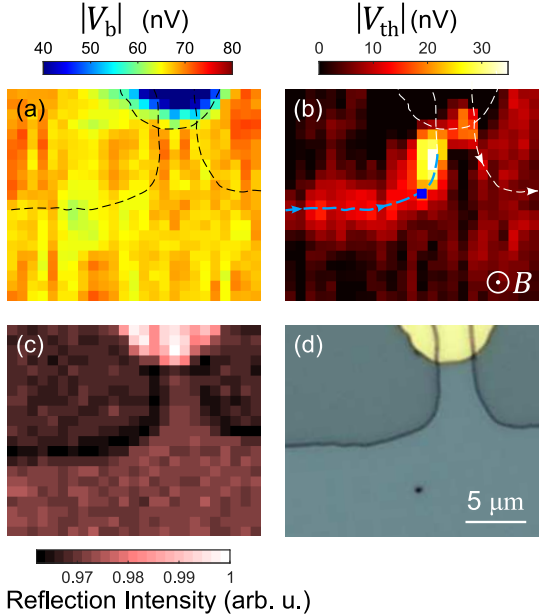}
\caption{(a) Spatial map of voltage measured across Ohmic electrodes (1) and (2) while the scanned laser power was $3.5$ nW, $V_{\text{fg}}$ was $-1.4$ V, and $T_{\text{L}}$ was 33 mK. (b) Spatial map of voltage measured across Ohmic electrodes (1) and (2) after subtracting a background component with the spatial distribution in (a). Laser power was $3.0$ nW, $V_{\text{fg}}=-0.5$ V, $T_{\text{L}}=1.7$ K, and $\nu=2$. Light blue dashed line and blue square correspond to measurements in Figs. \ref{fig.4} and \ref{fig.5} respectively. Arrows indicate the propagation direction of QH edge. (c) Spatial map of reflected laser intensity measured simultaneously with the measurement in (b). (d) Optical microscope image of the sample in the region mapped in (a) and (b). Outlines of the gate and etched 2DES boundary in the image are overlayed as dasshed lines on (a) and (b). $B=6.3$ T during the measurements ((a), (b) and (c)).}
\label{fig.3}
\end{figure}

We first investigated the spatial distribution of $V_{\text{th}}$ at $B=6.3$ T. We found that a voltage signal is detectable even when the laser illumination does not fall on the region of the sample containing the 2DES. This voltage signal, which we will call $V_{\text{b}}$, persists at 33 mK where we mapped it by scanning the laser in the region indicated by the dotted red box in Fig. \ref{fig.2}. This measurement of $V_{\text{b}}$ is shown in Fig. \ref{fig.3}(a). During this measurement we depleted the electrons under the front gate to  prevent any generation of $V_{\text{th}}$ across the gate. We see that $V_{\text{b}}$ has no notable spatial variation other than its diminution where the front gate blocks light from reaching the semiconductor. The voltage signal also appears spatially uniform at 1.7 K at locations far from the front gate (see Mapping Voltage Response to Illumination at the Ohmic Contacts in Supplementary Material for details). We conclude that a background signal independent of $V_{\text{th}}$ exists that is spatially uniform over the length scale that we map. We will later discuss the possible origin of $V_{\text{b}}$. We mapped $V_{\text{th}}$ at 1.7 K using a gate voltage $V_{\text{g}}=-0.5$ V. Figure \ref{fig.3}(b) shows this $V_{\text{th}}$ obtained after subtracting $V_{\text{b}}$ for this condition, which we assume to have the same spatial distribution as in Fig. \ref{fig.3}(a).

To interpret the map of $V_{\text{th}}$, we compare it to a map of reflected laser intensity measured simultaneously (Fig. \ref{fig.3}(c)). Matching this intensity map to the optical image of the sample in this region (Fig. \ref{fig.3}(d)), we located the etched boundaries of the QW and the gated region, which are shown as dotted outlines in Figs. \ref{fig.3}(a) and (b). It is apparent that $V_{\text{th}}$ is strongest when illumination is at the QH edge closest to the gate. And there is a decay with distance along the edge in the direction upstream from the gate. This observation can be understood from the chirality of heat transport in the QH edge channel. Heat injected upstream of the gate propagates clockwise and reaches the gate location, which creates the thermal gradient across the gate detected by $V_{\text{th}}$. Heat injected downstream of the gate only propagates away from the gate, so no thermal gradient is produced across the gate. The map of $V_{\text{th}}$ remained the same regardless of whether or not we applied source-drain current.

\begin{figure}
\onefigure[scale=0.35]{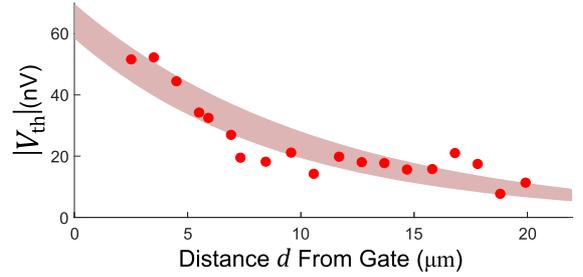} 
\caption{$V_{\text{th}}$ as a function of upstream distance from the gold gate along the etched 2DES boundary, obtained from the points overlayed by the blue dashed line in Fig. \ref{fig.3}(b). The fit of these data is $[64\pm6\;\text{nV}]\text{exp}(-d/[10\pm1\;\mu\text{m}])$, where the shaded region is the $68\%$ confidence interval.}
\label{fig.4}
\end{figure}

In Fig. \ref{fig.4} we plot $V_{\text{th}}$ as a function of distance $d$ from the gate along the edge upstream of the gate. The data plotted are taken from the points in Fig. \ref{fig.3}(b) overlayed by the blue dashed line which is a segment of the etching boundary visible in Fig. \ref{fig.3}(d). We fit these data to an exponential decay in order to extract a decay length of $10\pm1$ $\mu$m. We interpret this decay length as the distance over which the QH edge at elevated temperature cools into the environment at $T_{\text{L}}$. Our measured thermal decay length is shorter than the thermal decay length of $\sim20$ $\mu$m measured previously in the $\nu=1$ \cite{granger} and $\nu=2$ QH states \cite{nam} at a significantly lower temperature of 0.1 K.

\begin{figure}
\onefigure[scale=0.4]{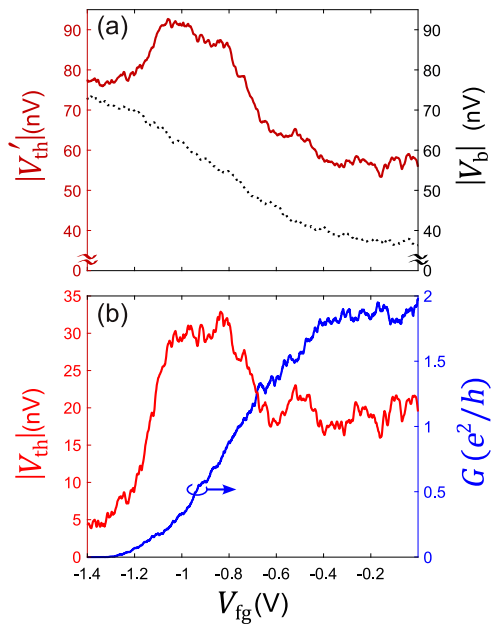} 
\caption{(a) Left axis: Voltage measured as a function of $V_{\text{fg}}$ while illuminating at the location of the blue square in Fig. \ref{fig.3}(b). Right axis: Average voltage measured as a function of $V_{\text{fg}}$ while illuminating at the four courners of the region mapped in Fig. \ref{fig.3}(b). Illuminating power was $3.5$ nW , $T_{\text{L}}=33$ mK, and $\nu=2$. (b) Left axis: The voltage obtained from the difference of the two curves in (a). Right axis: The four-terminal conductance of the constriction measured simultaneously with the voltage plotted on the right axis of (a).}
\label{fig.5}
\end{figure}

We next investigated how $V_{\text{th}}$ depends on $V_{\text{fg}}$. In Fig. \ref{fig.5}(a) we plot against $V_{\text{fg}}$ the voltage $V_{\text{th}}^{\prime}=V_{\text{th}}+V_{\text{b}}$ obtained before subtracting the background signal, which we measured while illuminating the 2DES edge at the location indicated by the blue square in Fig. \ref{fig.3}(b). Together we plot $V_{\text{b}}$ obtained by averaging the signal measured at the four corners of the region mapped in Fig. \ref{fig.3}; we assume that no significant thermoelectric signal was generated by illumination at these four corners because of their distance from the QH edge. Notably, $V_{\text{b}}$ exhibits a dependence on $V_{\text{fg}}$ which is opposite of $G$, plotted in Fig. \ref{fig.5}(b). A possible reason is that $V_{\text{b}}$ arises from the excitation of charged carriers in the GaAs substrate of the sample which modify the potential in the QW. When $G$ in the constriction is large, potential differences between Ohmic electrodes (1) and (2) can equilibrate, so $V_{\text{b}}$ is suppressed; whereas, when the constriction is depleted, no equilibration occurs and $V_{\text{b}}$ is maximum. 
In Fig. \ref{fig.5}(b) we plot $V_{\text{th}}$ obtained from the difference of the curves in Fig. \ref{fig.5}(a). $V_{\text{th}}$ has a peak where $G$ is in transition between zero and its quantized value of $e^2/h$. This is consistent with the work previously mentioned verifying the Mott relation in this type of device \cite{appleyard,nam}. The $\nu=1$ plateau in $G$ is absent, and the $\nu=2$ plateau is lower than the correct quantized value. These may be caused by backscattering across the constriction induced by potential fluctuations.

\begin{figure}
\onefigure[scale=0.33]{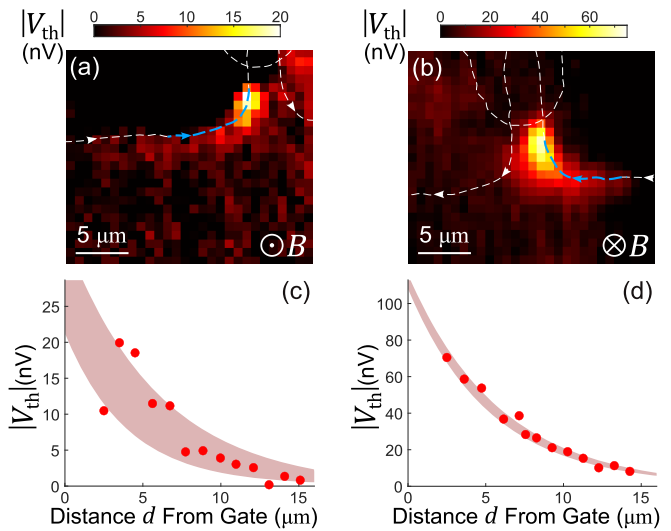} 
\caption{(a) and (b) Spatial maps of the voltage obtained as in Fig. \ref{fig.3}(b) while $B=3.6$ T was directed downward and upward respectively and $V_{\text{fg}}$ was $-458$ mV and $-520$ mV respectively. Illuminating power was $2$ nW, $T_{\text{L}}=1.7$ K, and $\nu=2$. Dotted lines are outlines of the gate and etched 2DES boundary. Arrows indicate the propagation direction of QH edge. (c) and (d) $V_{\text{th}}$ as a function of distance upstream from the gate obtained from points overlayed by the light blue dashed lines in (a) and (b) respectively. The fits of these data are (c) $[28\pm7\;\text{nV}]\text{exp}(-d/[5\pm1\;\mu\text{m}])$ and (d) $[112\pm5\;\text{nV}]\text{exp}(-d/[5.6\pm0.3\;\mu\text{m}])$, where the shaded regions are the $68\%$ confidence interval.}
\label{fig.6}
\end{figure}

Next, we repeated the mapping of $V_{\text{th}}$, still at $\nu=2$, while applying a weaker field of $B=3.6$ T directed both upward and downward perpendicular to the 2DES (Figs. \ref{fig.6}(a) and (b) respectively). At this weaker $B$, $V_{\text{b}}$ became undetectably small. The difference in the magnitude of $V_{\text{th}}$ between the two plots is due to the different conditions $V_{\text{fg}}$ of the gate during the measurements. For both $B$ directions, the $V_{\text{th}}$ signal shows the same behavior seen at 6.3 T, which is consistent with chiral downstream heat propagation in the QH edge. As in Fig. \ref{fig.4}, in Figs. \ref{fig.6}(c) and (d) we plot $V_{\text{th}}$ as a function of upstream distance $d$ from the gate. Data are taken along the blue dashed lines in Figs. \ref{fig.6}(a) and (b) respectively, and fit to extract thermal decay lengths of $5\pm1$ $\mu$m and $5.6\pm0.3$ $\mu$m respectively. The measured thermal decay length has become shorter, possibly caused by the decrease in $B$ or the decrease in $n_{\text{e}}$ throughout the device by the same proportion. A more systematic study of how thermal decay length depends on experimental conditions is needed to understand the mechanism of heat dissipation from the QH edge. 

We will now discuss possible challenges to our interpretation of the measured $V_{\text{th}}$ signal as thermoelectric voltage. Photovoltage is a phenomenon which requires consideration in our experiment.  Photovoltage generated by microscopic laser photoexcitation of electrons into the QH edge was measured in single-heterojunction QWs \cite{shashkin,vanzalinge}. In that work, spatial maps of photovoltage with peak magnitudes of $\sim200$ nV and $\sim1$ $\mu$V were detected for illumination powers of $\sim1$ nW and $\sim1$ $\mu$W respectively. The photovoltage occurred in the absence of any gate or constriction of the 2DES and had a localized response to illumination at the 2DES edge. In Ref. \cite{vanzalinge}, the photovoltage could be seen to decay with distance from the Ohmic contacts used for detection with a decay length of $\sim1000$ $\mu$m at 1.4 K. 

We mapped the voltage response to illumination while scanning the laser at the Ohmic contacts because the photovoltage should be maximized there (see Mapping Voltage Response to Illumination at the Ohmic Contacts in Supplementary Material for details). However, we were unable to detect any signal with the characteristics of photovoltage. We observed only the spatially uniform background voltage discussed above, so we conclude that any photovoltage that may be generated is too small to be detected using the illumination power of our experiment. This is probably because the QW in our device is square in the growth direction, which causes photoexcited holes to remain in the QW rather than vertically drifting out of the well as in previous experiments. The presence of photoholes at the edge of the 2DES, which have the same chirality as electrons, can result in a reduction or complete negation of the photoexcited charge.

It is also important to consider the thermalization of photoexcited carriers in our experiment. Here, "thermalization" entails the processes in which electrons occupying energy states above the Fermi distribution relax into the Fermi distribution while raising the Fermi temperature. The QH edge is a one-dimensional Tomonaga-Luttinger liquid \cite{hashisaka,jompol} which, in an ideally clean system at sufficiently low excitation energy, lacks the thermalization processes that are present in Fermi liquids \cite{giamarchi}. Confirmation of this was provided by quantum-dot energy spectroscopy measurements identifying a nonequilibrium energy distribution as far away as $15$ $\mu$m from the point of electrical excitation of the edge state \cite{itoh}. However, the QH edge in that work was defined by electrostatic gating, which creates a smooth confinement potential. It was shown that when electrons having several times the longitudinal optical (LO) phonon energy (36 meV) were injected into the QH edge, electrostatically broadening the edge potential suppressed scattering with these LO phonons, which hindered efficient equilibration \cite{akiyama}. Further results, also using excitation greater than the LO phonon energy, indicated that relaxation by electron-electron scattering is also suppressed by edge potential broadening and increased excitation energy \cite{ota}. 

The photoexcited electron-hole pairs in our experiment have an initial combined excitation energy of $\sim45$ meV, meaning that LO phonon scattering is important for thermalization. However, because the QH edge in our device is defined by etching, the confinement potential should be relatively sharp compared to a gate-defined potential \cite{venkatachalam}. It is conceivable, therefore, that the LO phonon scattering length is among the lowest values measured in Ref. \cite{ota}, i.e. $<1$ $\mu$m. This scattering length is a lower bound for the thermalization length, i.e. the distance over which thermalization occurs, in our experiment. If the thermalization length is larger than the laser spot size, the heating region will be larger than we initially assumed. Considering this possibility, the thermal decay lengths we measure should be interpreted as an upper bound. Lastly, we note that because the constriction width is comparable to the measured thermal decay lengths, details of the potential landscape inside the constriction can significantly impact the effective distance between temperature probe and heating point. To address this limitation, a quantum point contact gate structure could be used as the temperature sensor.

In conclusion, we have demonstrated measurement of the thermal decay length of the QH edge by real-space optical mapping of thermovoltage generated at a local thermometer. We find the thermal decay length to change with experimental parameters, which calls for further investigation of the mechanism of energy dissipation from the QH edge.

\acknowledgments
The authors are grateful to T. Fujisawa for fruitful discussions. This work was supported by Grants-in-Aid for Scientific Research (Grant Nos. 21F21016, 19H05603, 21J14386 and 21H05188) from the Ministry of Education, Culture,
Sports, Science, and Technology (MEXT). J. N. M. and A. K. were supported by Grants-in-Aid from MEXT.

\textit{Data availability statement}: The data that support the findings of this study are available upon reasonable request
from the author.

\end{document}